\documentclass[%
aip,
cp,  % Conference Proceedings
amsmath,amssymb,%nobibnotes,
preprint,%
]{revtex4-1}

\usepackage{graphicx}% Include figure files
\usepackage{dcolumn}% Align table columns on decimal point
\usepackage{bm}% bold math
\usepackage[utf8]{inputenc}
\usepackage[T1]{fontenc}
\usepackage{mathptmx} 

\begin{document}

\title{Beam Asymmetries ($\Sigma$)  for $\pi^0$, $\eta$ and $\eta'$ in Photoproduction at GlueX}% Force line breaks with \\

\author{William McGinley} % Write as First name Surname
 \email[Corresponding author: ]{wmcginle@andrew.cmu.edu}
%\author{Curtis A. Meyer}
% \email{cmeyer@cmu.edu}
\affiliation{Carnegie Mellon University, 5000 Forbes Ave, Pittsburgh, PA 15213, USA}
\author{Tegan Beattie}%
  \email{beattite@uregina.ca}
\affiliation{
University of Regina, 3737 Wascana Pkwy, Regina, SK S4S 0A2, Canada% Force line breaks with \\ if necessary
}

\collaboration{on behalf of the GlueX Collaboration}

\date{\today} % It is always \today, today, but any date may be explicitly specified
              % Not printed for conference proceedings

\begin{abstract}
The GlueX experiment is a photoproduction experiment located at Thomas Jefferson National Lab in Newport News, Virginia. GlueX is capable of making beam asymmetry ($\Sigma$) measurements using a tagged, linearly-polarized 9 GeV photon beam incident on a hydrogen target. Measurements of the beam asymmetry for the exclusive reactions $\gamma p \rightarrow \pi^0 p$, $\gamma p \rightarrow \eta p$ and $\gamma p \rightarrow \eta' p$ will provide insight into the meson production mechanisms. GlueX measurements are the first beam asymmetry results for the $\eta$ and $\eta’$ in this energy range and are expected to further constrain Regge theory models for photoproduced pseudoscalar mesons. This talk will present preliminary results of the photon beam asymmetries as a function of the Mandelstam variable, t, for multiple $\eta$ decay modes and the $\eta'\rightarrow\pi^+\pi^-\eta$ decay mode.
\end{abstract}

\maketitle

\section{\label{sec:into}INTRODUCTION\protect\\}

\indent The primary goal of GlueX is to map out a spectrum of hybrid exotic mesons. An important early step to accomplish this goal is to measure observables, which are the angular distributions of final state particles. The $\Sigma$ beam asymmetry is one such observable that provides insight into the production mechanism of pseudoscalar mesons. The production of pseudoscalar mesons at 9 GeV is expected to be described by the Regge Model~\cite{BARKER1975347,PhysRevD.92.074004}. The photoproduction of $\pi^0$, $\eta$ and $\eta'$ involves the t-channel exchange of Reggeons with allowed quantum numbers $J^{PC}=1^{--}$ or $J^{PC}=1^{+-}$~\cite{PhysRevD.95.034014}. The measurement of $\Sigma$ provides insight into the contributions of the natural exchange, $J^{PC}=1^{--}$, and the unnatural exchange, $J^{PC}=1^{+-}$, to the production mechanism.  While $\Sigma_{\eta}$ and $\Sigma_{\eta'}$ provide valuable information on their own, the ratio of the two can shed light on the contributions of hidden strangeness exchange ($s\bar{s}$) states such as the $\phi$ and $h'$ and axial vector meson ($b$ and $h$) exchange~\cite{2017362}.

\indent There have been several experiments to make high precision measurements of $\Sigma_{\eta}$~\cite{Vartapetian:1980cn, PhysRevLett.81.1797, Elsner2007, Bartalini2007, PhysRevC.78.015203, COLLINS2017213} and a limited set of measurements of $\Sigma_{\eta'}$~\cite{COLLINS2017213,LeviSandri2015} at beam energies of less than 2 GeV . In this energy regime $\Sigma$ provides insight into the nucleon resonances. Experiments have measured $\Sigma$ and the other 15 pseudoscalar meson polarization observables to apply constraints to the models of the helicity amplitudes of the excited nucleon states. GlueX runs with a beam energy of 9 GeV and is therefore insensitive to the nucleon resonance spectrum. However, the t-channel component of the model extends from the low energy regime up to the GlueX energy. A precise measurement of $\Sigma$ by GlueX could apply new constraints to the t-channel component of the model and be extrapolated back to the low energy regime to help understand the nucleon resonance spectrum. 

\indent The only previous measurement at high-energy of $\Sigma_{\eta}$ was made by GlueX~\cite{PhysRevC.95.042201} for only the $\eta\rightarrow2\gamma$ mode. The analysis presented here improves the precision of the same decay mode and also utilizes the $\eta\rightarrow\pi^+\pi^-\pi^0$ and $\eta\rightarrow3\pi^0$ decay modes to improve the precision on measuring $\Sigma$ for the $\gamma p \rightarrow \eta p$ reaction. The first high-energy measurement of $\Sigma$ for the $\gamma p \rightarrow \eta' p$ reaction using the $\eta'\rightarrow \pi^+\pi^-\eta$ decay is also presented here. The data that are presented represents a total of 20.8$pb^{-1}$ and were collected at Thomas Jefferson National Lab by the GlueX experiment. More details of this analysis were recently submitted for publication~\cite{Adhikari:2019gfa}.

\section{\label{sec:GXDetector}GlueX DETECTOR}

\begin{figure}
\includegraphics[width=0.75\textwidth]{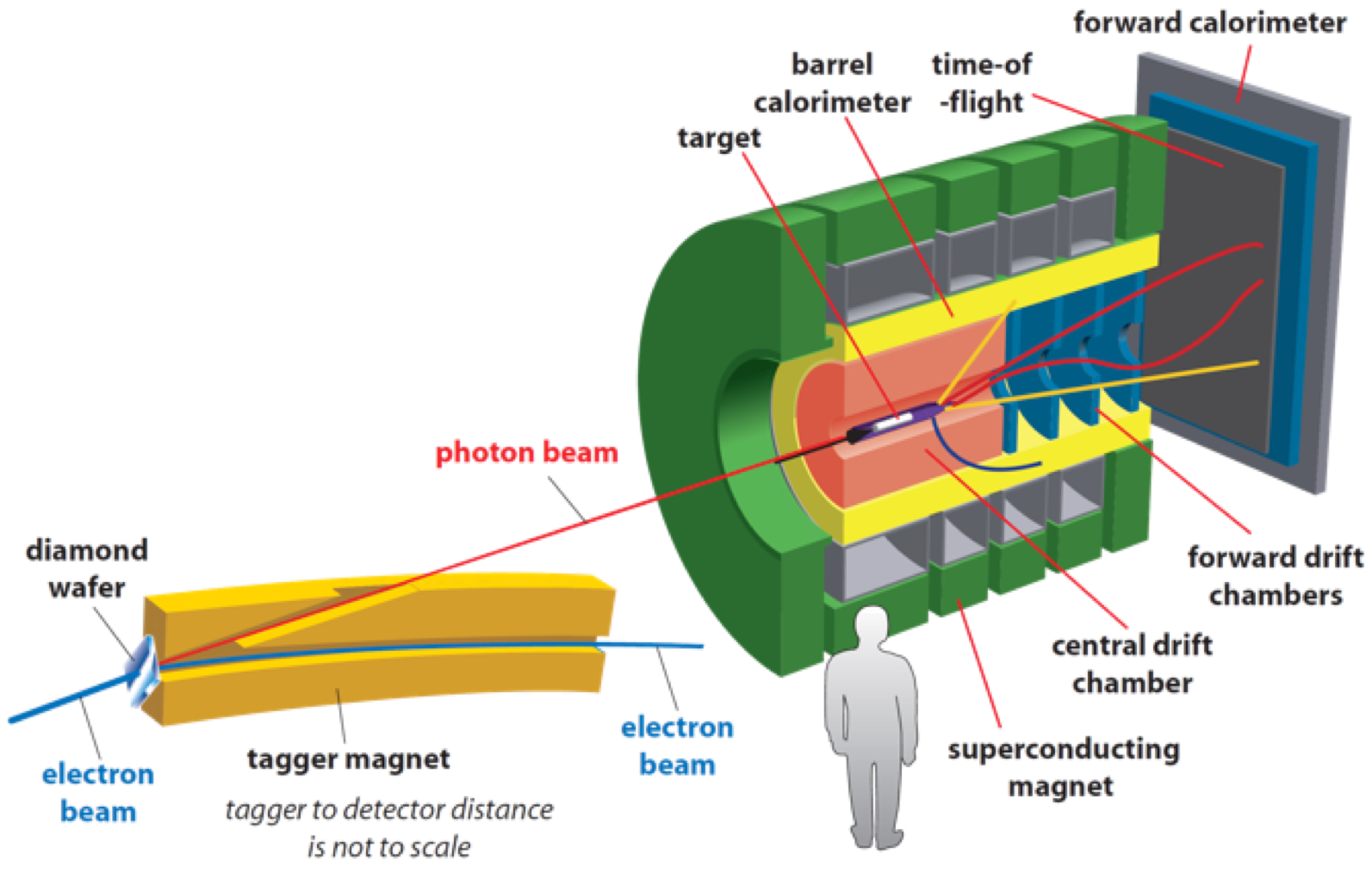}
\caption{\label{fig:gluex_detector} GlueX Detector}
\end{figure}

The GlueX spectrometer is a nearly 4$\pi$ hermetic detector that is capable of detecting neutral and charged particles along with performing robust particle identification. The Continuous Electron Beam Accelerator Facility (CEBAF) provides an electron beam that is incident on a thin diamond radiator and converted to a 9 GeV linearly-polarized photon beam via Bremsstrahlung radiation. The photon beam is measured to have an average polarization of around 35\% in the coherent peak region by a Triplet Polarimeter~\cite{Dugger:2017zoq}. The scattered electrons have their energy tagged in order to determine the energy of the photon beam allowing for exclusive reconstruction of events. The general organization of the many components of the GlueX spectrometer can be seen in Fig.~\ref{fig:gluex_detector}. Surrounding the target is a Start Counter~\cite{Pooser:2019rhu} used for particle identification (PID). The Central Drift Chamber envelops the Start Counter; immediately downstream of the Central Drift Chamber is a Forward Drift Chamber ~\cite{Pentchev:2017omk}. The drift chambers provide charged particle tracking. They are surrounded by a barrel electromagnetic calorimeter~\cite{Beattie:2018xsk}, which, in conjunction with the planar Forward Calorimeter, is responsible for detecting neutral particles. In front of the Forward Calorimeter is a Time-of-Flight detector responsible for PID measurements. The Barrel Calorimeter sits inside a solenoid magnet that applies a 2T magnetic field along the direction of the incoming beam.

\indent The data were collected with four different orientations of the diamond comprising two sets of data containing two orientations each. Two of the orientations set the linear polarization plane parallel and perpendicular to the floor ($x-z$ plane), referred to as PARA and PERP, respectively. These two orthogonal orientations comprise one set of data. The other set of data introduces a +45$^{\circ}$ azimuthal offset to both the PARA and PERP orientations. This results in two independent sets of data each with orthogonal directions of polarization. The orthogonality of the polarization directions simplifies the extraction of the beam asymmetry, $\Sigma$. The two independent sets of data allows for systematic studies between the two. 

\section{\label{sec:method}Method}

\begin{figure}
\includegraphics[width=0.8\textwidth]{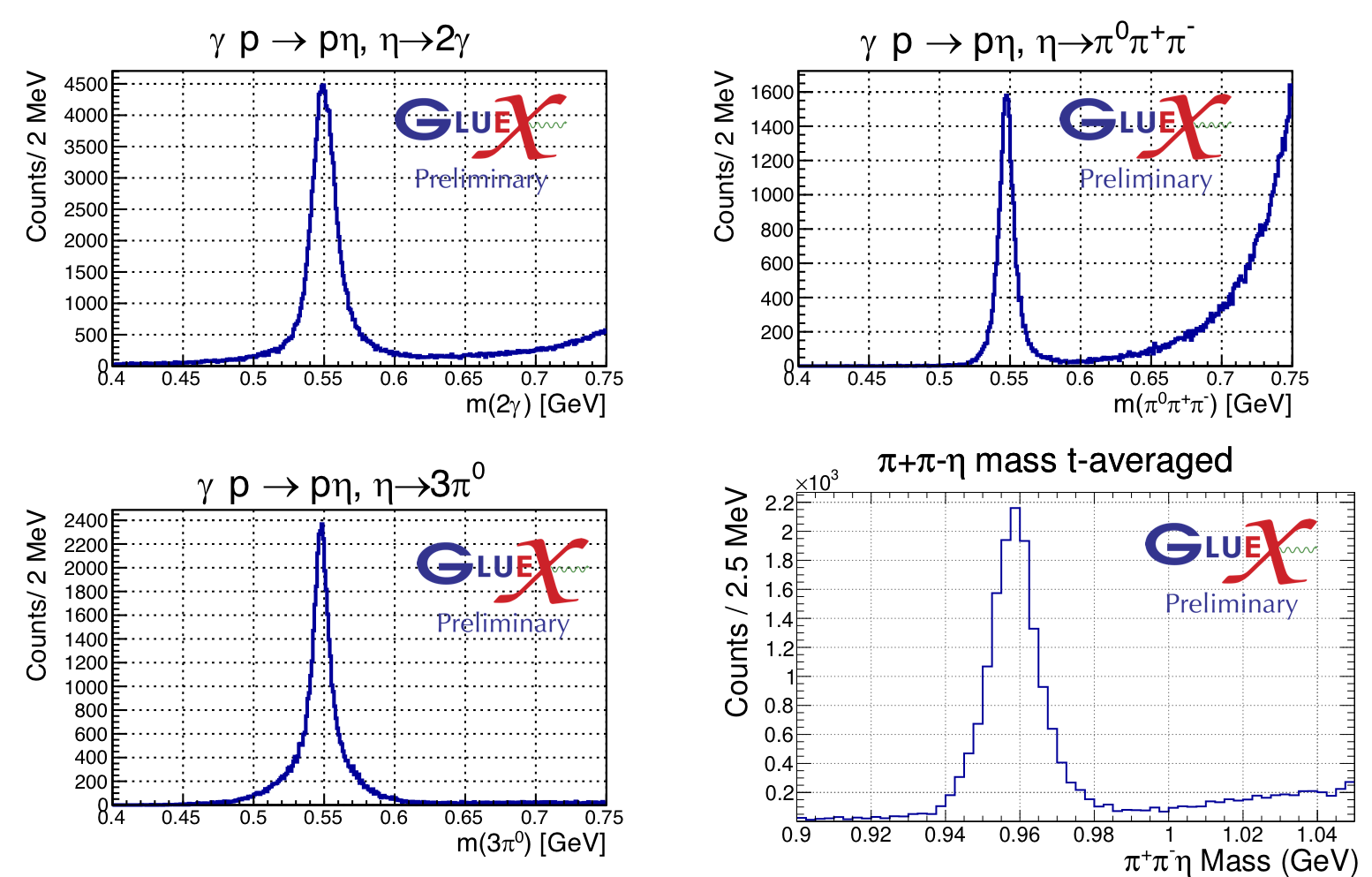}
\caption{\label{fig:masses} Invariant mass distributions after the event selection criteria are applied.}
\end{figure}

\indent In order to extract $\Sigma$ a clean sample is needed, the invariant mass distributions are shown in Fig.~\ref{fig:masses}. These are produced by reconstructing exclusive events with a kinematic fit performed. The kinematic fit conserves energy and momentum, constrains the event vertex and constrains the mass of of the $\pi^0$ ($\eta$) in the $\eta\rightarrow\pi^+\pi^-\pi^0$ ($\eta'\rightarrow\pi^+\pi^-\eta$) decays.

\begin{figure}
\includegraphics[width=0.6\textwidth]{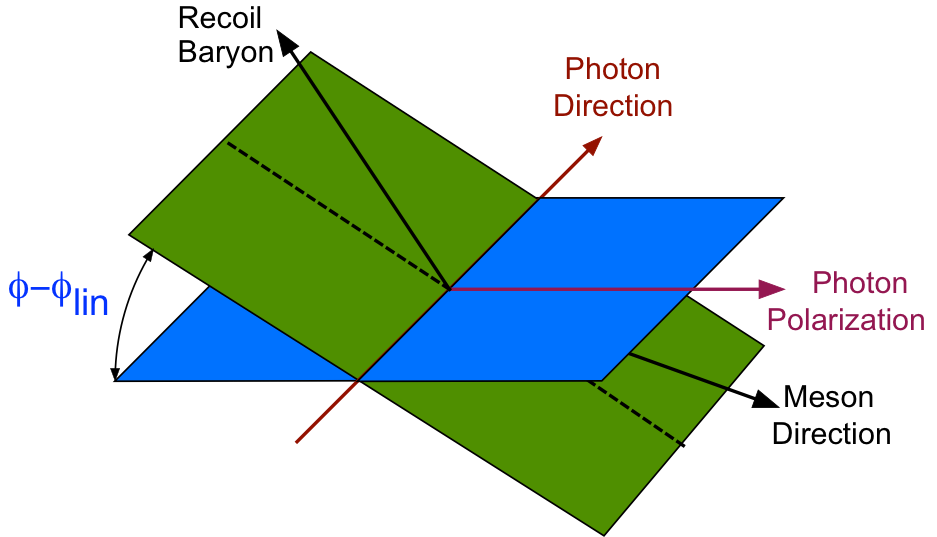}
\caption{\label{fig:asym_diag} Illustration to display how the relevant angles are defined in the LAB frame.}
\end{figure}

\indent $\Sigma$ is extracted experimentally by measuring the angle of the reaction plane, $\phi$, while keeping the polarization angle, $\phi_\gamma$, fixed, see Fig.~\ref{fig:asym_diag}. When the reaction plane is perpendicular to the polarization plane the natural parity exchange will contribute positively to $\Sigma$. When the reaction plane is parallel to the polarization plane the unnatural parity exchange will contribute negatively to $\Sigma$, as shown in Eq.~\ref{eqn:Sigma_def}. If the pseudoscalar meson production mechanism is dominated by the natural exchange then $\Sigma$ will be consistent with +1; if the production mechanism is dominated by the unnatural exchange then $\Sigma$ will be consistent with -1. 

\indent For the photoproduction of pseudoscalar mesons with a linearly polarized photon beam and an unpolarized target, the polarized cross section is given by~\cite{BARKER1975347}

\begin{equation}
\sigma_{pol}(\phi, \phi^{lin}_{\gamma}) = \sigma_{unpol} \left[ 1 - P_\gamma\Sigma \cos\left(2(\phi - \phi^{lin}_{\gamma})\right)\right],
\end{equation}
where $\sigma_{unpol}$ is the unpolarized cross section, $P_\gamma$ is the magnitude of the photon beam polarization, $\phi$ is the azimuthal angle of the production plane, $\phi^{lin}_{\gamma}$ is the azimuthal angle of the photon beam linear polarization plane and $\Sigma$ is the beam asymmetry observable of interest. Therefore the $\Sigma$ beam asymmetry can be constructed as

\begin{equation}
\Sigma = \frac{\sigma_\perp - \sigma_\parallel}{\sigma_\perp + \sigma_\parallel}.
\label{eqn:Sigma_def}
\end{equation}

\noindent where

\begin{align}
\label{eqn:sigma_para}
\sigma_\parallel(\phi) &= \sigma_{pol}(\phi, \phi^{lin}_{\gamma}=0) = \sigma_{unpol} (1 - P_\parallel \Sigma \cos2\phi)\\
\label{eqn:sigma_perp}
\sigma_\perp(\phi) &= \sigma_{pol}(\phi, \phi^{lin}_{\gamma}=90) = \sigma_{unpol} (1 + P_\perp \Sigma \cos2\phi)
\end{align}

\noindent and $P_\parallel$ and $P_\perp$ are the magnitude of the photon beam polarization in the PARA and PERP orientations, respectively.

The GlueX detector is designed to be symmetric in $\phi$ and thus have a uniform acceptance and efficiency, but here we consider the general case of some arbitrary $\phi$-dependent detector acceptance and define the methods for extracting $\Sigma$ which cancel this detector acceptance. In the absence of background, one can define a polarization-dependent yield asymmetry, $S(\phi)$ as 

\begin{equation}
S(\phi) = \frac{Y_\perp(\phi) - F_R Y_\parallel(\phi)}{Y_\perp(\phi) + F_R Y_\parallel(\phi)} = \frac{\sigma(\phi)_\perp A(\phi) - \sigma(\phi)_\parallel A(\phi)}{\sigma(\phi)_\perp A(\phi) + \sigma(\phi)_\parallel A(\phi)}
\label{eqn:asym}
\end{equation}

\noindent where $F_R$ is the flux ratio between $Y_{\perp}$ and $Y_{\parallel}$. The measured yield asymmetry $S(\phi)$ can be fit to a function of the following form:

\begin{equation}
%S(\phi) = 
\frac{Y(\phi)_\perp - F_RY(\phi)_\parallel}{Y(\phi)_\perp + F_RY(\phi)_\parallel}  = \frac{(P_\perp + P_\parallel) \Sigma \cos2(\phi-\phi_0)}{2 + (P_\perp - P_\parallel) \Sigma \cos2(\phi-\phi_0)}. %P \Sigma \cos2\phi.
\label{eqn:asym_fit}
\end{equation}

\begin{figure}
\includegraphics[width=0.99\textwidth]{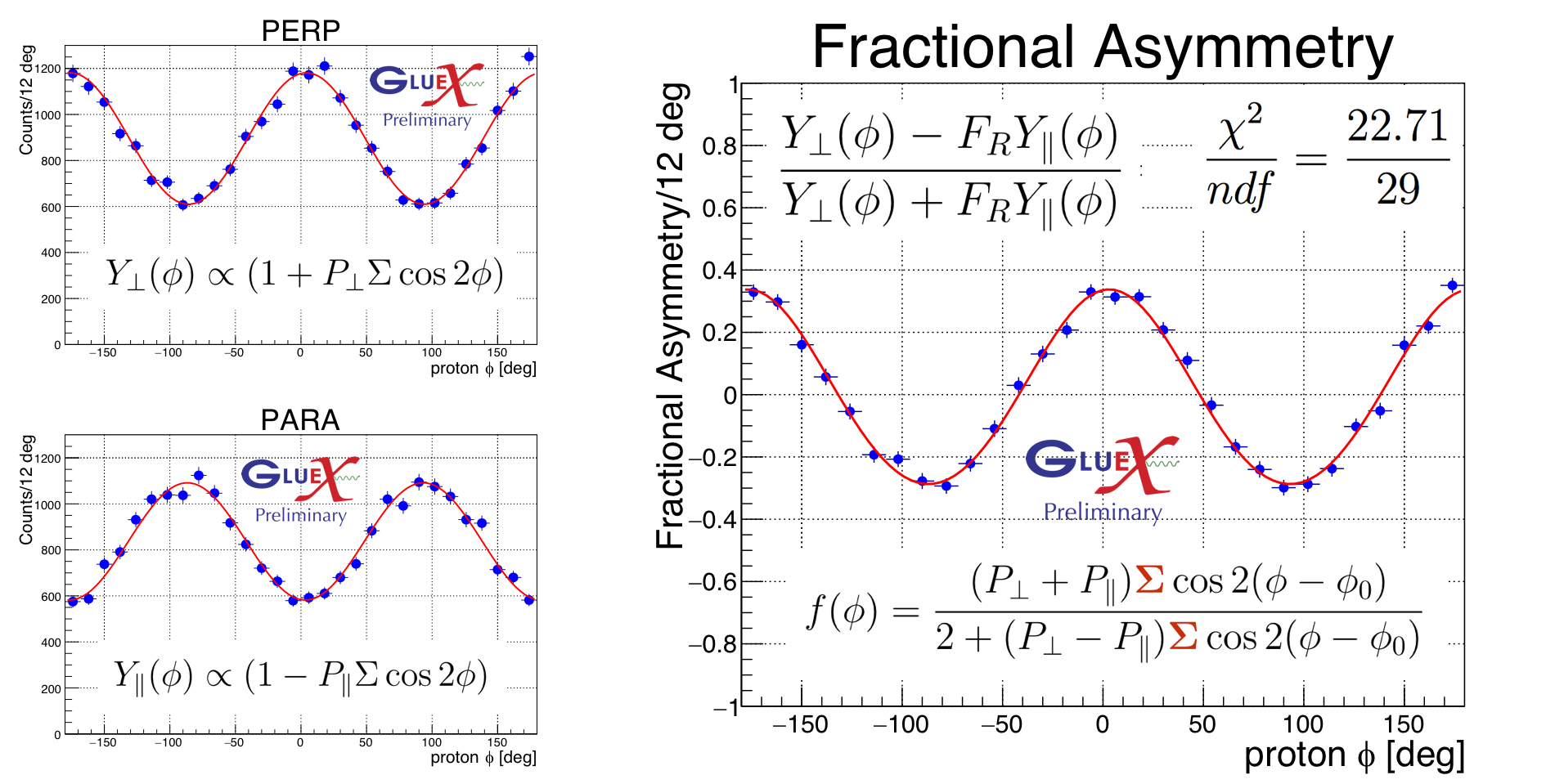}
\caption{\label{fig:fit_ex} Top left shows the azimuthal yield with the diamond in the PERP (90$^{\circ}$) orientation. The lower left shows the azimuthal yield with the diamond in the PARA (0$^{\circ}$) orientation. Top right shows the fractional asymmetry distribution from Eq.~\ref{eqn:asym} with the fit to extract $\Sigma$ using Eq.~\ref{eqn:asym_fit}.}
\end{figure}

\noindent Therefore with independent measurements of the flux, the polarization plane offsets ($\phi_0$) and polarization magnitudes for PARA and PERP datasets, one can determine $\Sigma$ independently of any $\phi$-dependent detector acceptance effects. $\Sigma$ remains as the only free parameter in the fit and is extracted directly. An example distribution and fit to extract $\Sigma$ for the $\eta\rightarrow2\gamma$ mode can be seen in Fig.~\ref{fig:fit_ex}.

\indent There is a significant background contribution in the signal mass region for the $\eta\rightarrow2\gamma$,  $\eta\rightarrow\pi^{+}\pi^{-}\pi^{0}$ and $\eta'\rightarrow\pi^{+}\pi^{-}\eta$ reactions. An asymmetry measurement due to the background will generally have a different value than the asymmetry of the signal process and could affect the value of the beam asymmetry for the process of interest. The signal asymmetry can be corrected for the background contribution using the following expression

\begin{equation}
\Sigma_{signal} = \frac{\Sigma_{peak} - f\Sigma_{SB}}{1-f}
\end{equation}

\noindent where $f$ is the fraction of background events in the signal mass window, $\Sigma_{peak}$ is the measured asymmetry in the signal mass window and $\Sigma_{SB}$ is the background asymmetry from a sideband region.  $\Sigma_{SB}$ is measured by selecting events with higher mass than the signal region where the background is present and fitting the asymmetry using the same method described above. The observed background asymmetry from the sideband regions ends up being opposite in sign to the peak asymmetry for these channels. %As shown in Fig.~\ref{fig:cutSummary_2g} and ~\ref{fig:cutSummary_3piq}, after all selection cuts are applied some background events remain above the $\eta$ mass range for the $\eta\rightarrow2\gamma$ and $\eta\rightarrow\pi^{+}\pi^{-}\pi^{0}$ decays.  These events are the result of the much larger cross section of the $\gamma p \rightarrow \omega p$ reaction. The background in the $\eta\rightarrow2\gamma$ is due to the $\omega$ radiative decay to $\pi^0\gamma$ where one of the photons from the $\pi^{0} \rightarrow \gamma\gamma$ decay is undetected. The background in the $\eta\rightarrow\pi^{+}\pi^{-}\pi^{0}$ is due to the  $\omega\rightarrow\pi^{+}\pi^{-}\pi^{0}$ hadronic decay. The contribution of these events is understood by generating a sample of $\omega$ MC events generated by two omega decay generators for each of these two decay modes.

\section{Results}

The results of the measured $\Sigma$ as a function of momentum transfer, -t, are compared with the previous GlueX results and several theoretical predictions. The data points are positioned at the mean value of the t-distribution in each bin and the length of each horizontal error bar is the rms of the t-distribution in that bin. The height of each vertical error bar is equal to the total error while the height of the shaded box is equal to the systematic uncertainty associated with the measurement. The width of the shaded boxes is arbitrary for illustrative purposes.  

\begin{figure}[!h]
\includegraphics[width=0.85\textwidth]{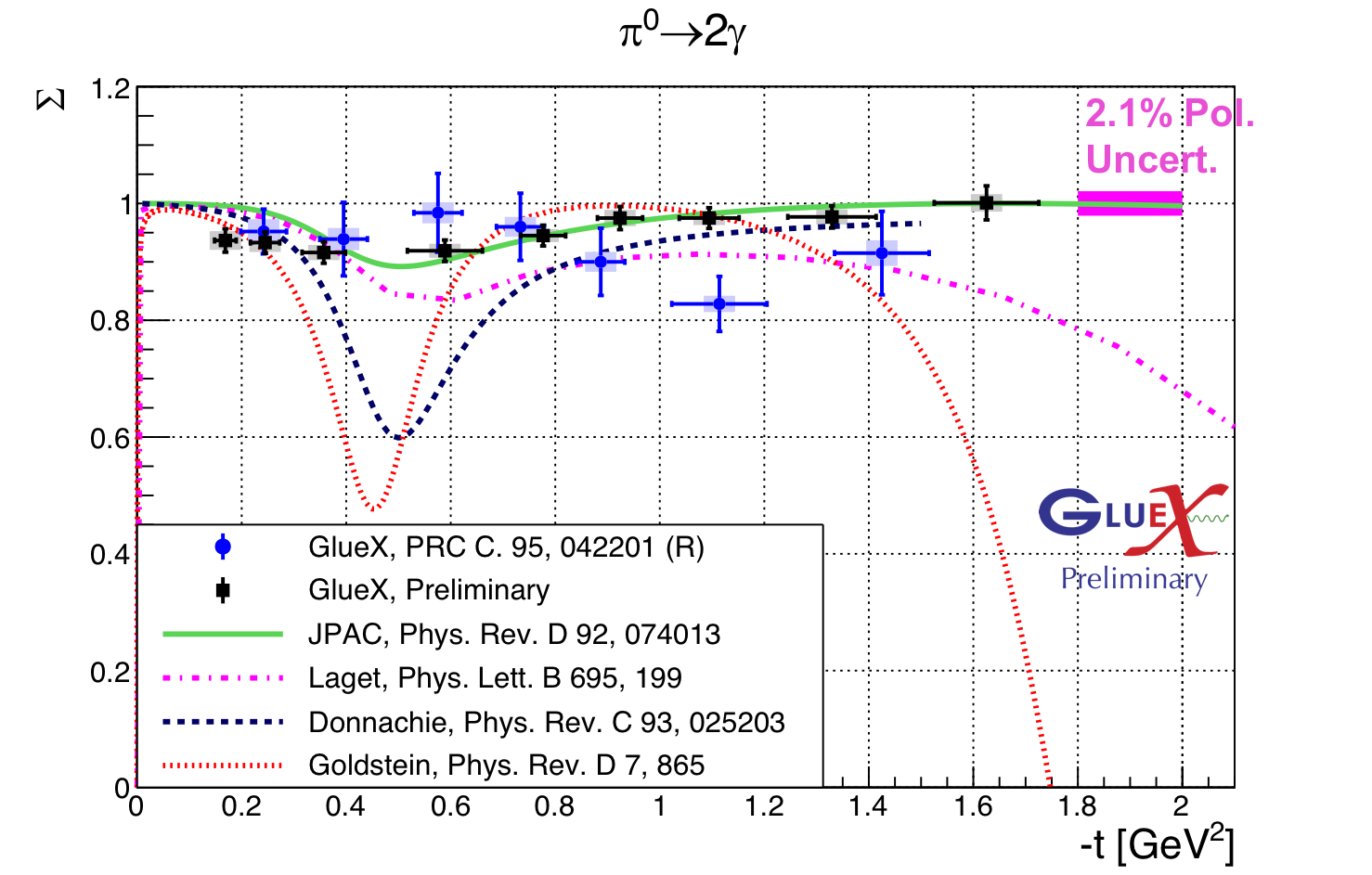}
\caption{\label{fig:pi0_asym} Measured beam asymmetry, $\Sigma$, as a function of -t for the $\gamma p \rightarrow \pi^0 p, \pi^0 \rightarrow 2\gamma$ reaction. There is a relative uncertainty of 2.1\% due to the polarization measurement.}
\end{figure}

\begin{figure}[!h]
\includegraphics[width=0.85\textwidth]{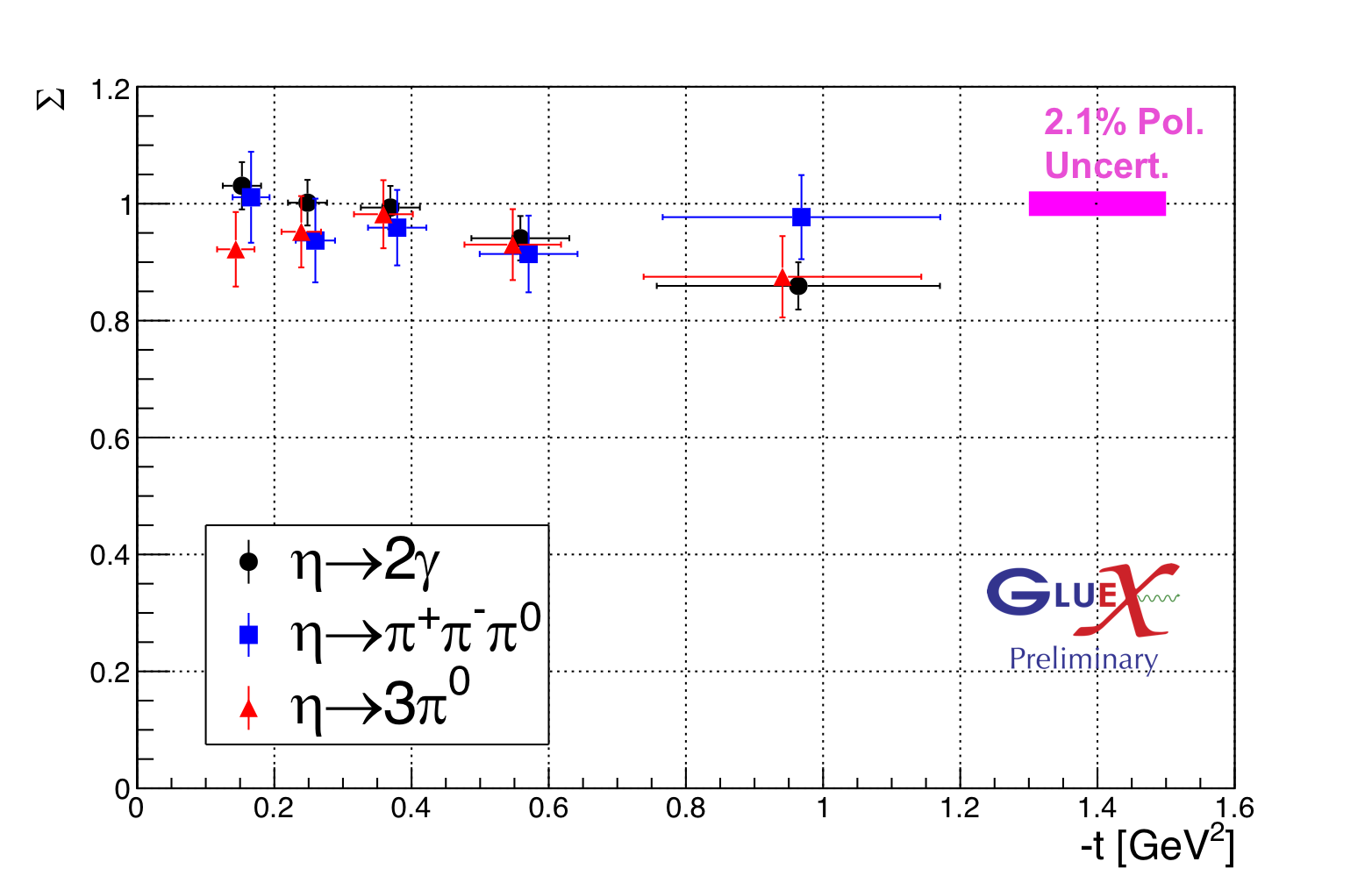}
\caption{\label{fig:pi0_asym} Measured beam asymmetry, $\Sigma$, as a function of -t for the $\gamma p \rightarrow \eta p$ reaction for the three most dominant decay modes of $\eta$. There is a relative uncertainty of 2.1\% due to the polarization measurement.}
\end{figure}

\begin{figure}[!h]
\includegraphics[width=0.85\textwidth]{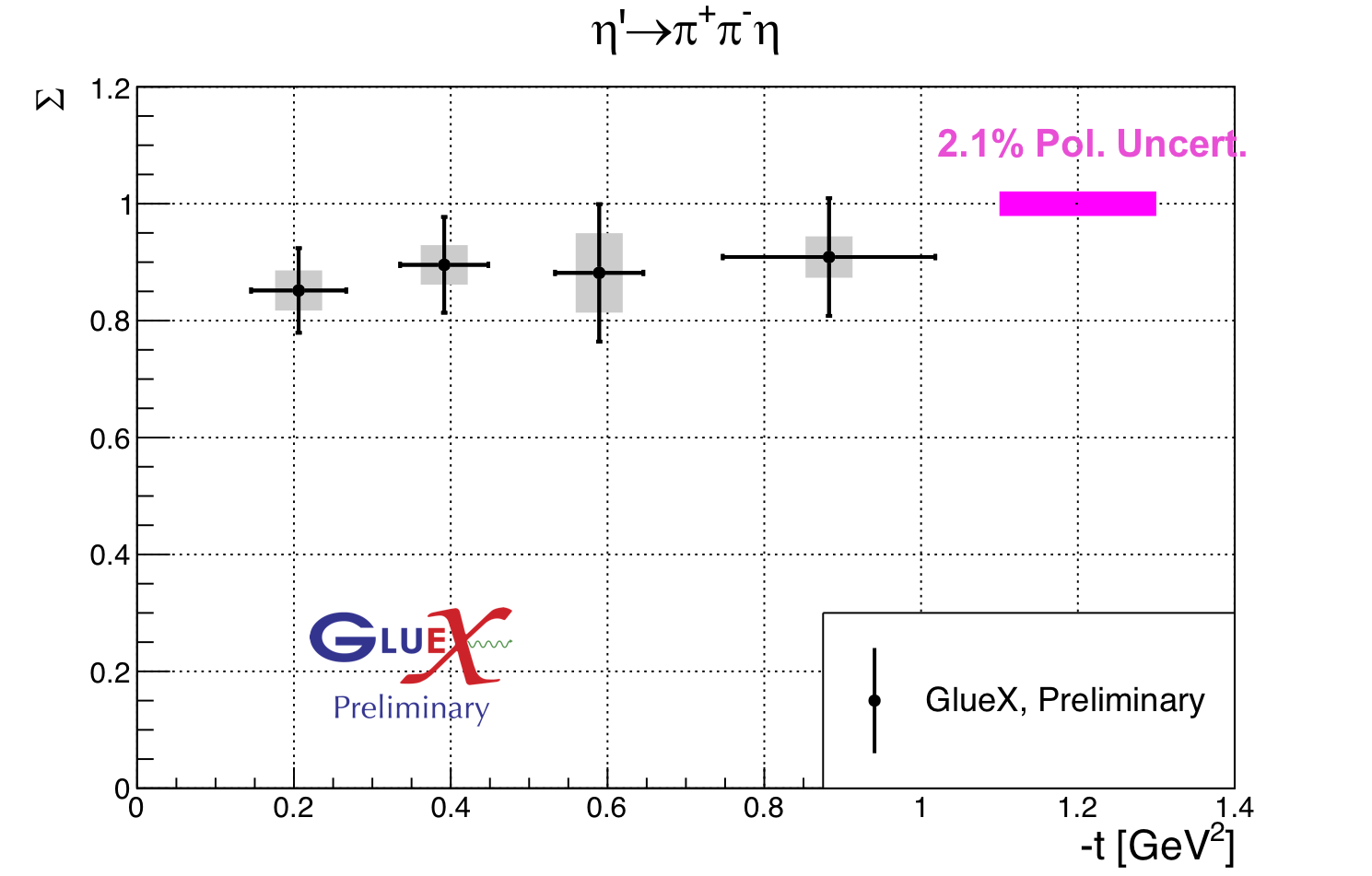}
\caption{\label{fig:pi0_asym} Measured beam asymmetry, $\Sigma$, as a function of -t for the $\gamma p \rightarrow \eta' p, \eta' \rightarrow \pi^+\pi^-\eta, \eta \rightarrow 2\gamma$ reaction. There is a relative uncertainty of 2.1\% due to the polarization measurement.}
\end{figure}

\begin{figure}[!h]
\includegraphics[width=0.85\textwidth]{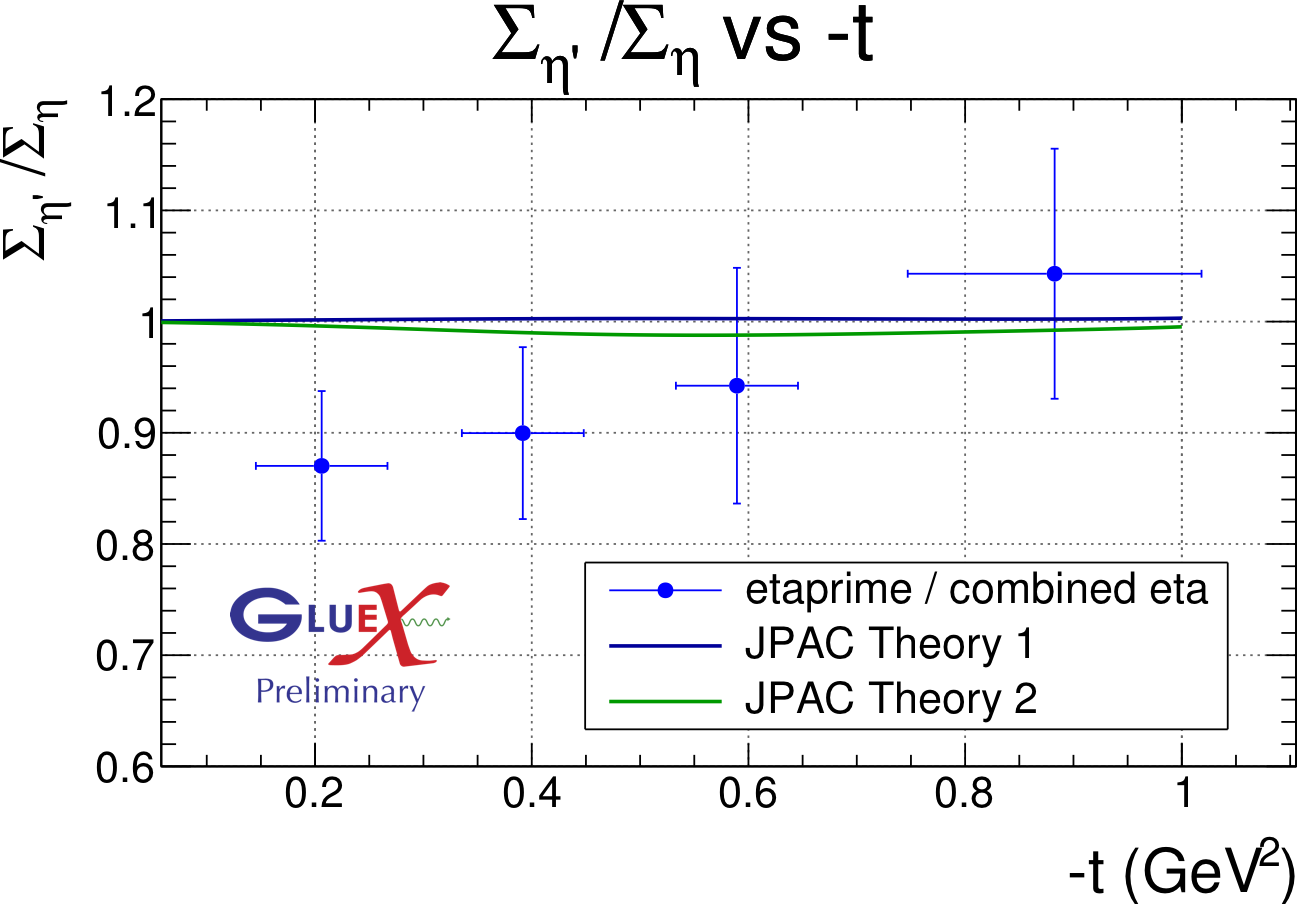}
\caption{\label{fig:pi0_asym} Measured beam asymmetry ratio between $\eta$ and $\eta'$ as a function of -t.}
\end{figure}

\section{Conclusion}

For each of the three reactions: $\gamma p \rightarrow \pi^0 p$, $\gamma p \rightarrow \eta p$ and $\gamma p \rightarrow \eta' p$, the value of $\Sigma$ is close to unity across the measured t-range, indicating that both production mechanisms are dominated by natural exchange, $J^{PC} = 1^{--}$. These are the first measurements of $\Sigma$ in this energy range for $\gamma p \rightarrow \eta' p$ and for multiple decay modes of $\eta$ in $\gamma p \rightarrow \eta p$. The ratios of these asymmetries are consistent with the predictions made by JPAC. This analysis will be continued with the full data set from the GlueX Phase 1 running which will increase the statistics by a factor of 4. The cross-sections for both reactions are currently being analyzed. 

\begin{acknowledgments}
The work of the Medium Energy Physics group at Carnegie Mellon University was supported by DOE Grant No. DE-FG02-87ER40315. This material is based upon work supported by the U.S. Department of Energy, Office of Science, Office of Nuclear Physics under contract DE-AC05-06OR23177.
\end{acknowledgments}

\nocite{*}
\bibliography{McGinley-MENU}% Produces the bibliography via BibTeX.

\end{document}